\documentstyle[preprint,aps,prabib]{revtex}
\begin{document}
\draft
\title{Confining properties of the classical\\
 $SU(3)$ Yang - Mills theory}
\author{V. Dzhunushaliev
\thanks{E-mail address: dzhun@freenet.bishkek.su}}
\address{Theoretical Physics Department,\\
the Kyrgyz State National University,\\ 
720024, Bishkek, Kyrgyzstan}
\date{}
\maketitle
\begin{abstract}
The spherically and cylindrically symmetric solutions of the $SU(3)$
Yang - Mills theory are obtained. The corresponding gauge potential
has the confining properties. It is supposed that: a) the spherically
symmetric solution is a field distribution of the classical ``quark'' 
and in this  sense it is similar to the Coulomb potential; 
b) the cylindrically symmetric 
solution describes a classical field ``string'' (flux tube) between 
two ``quarks''. It is noticed that these solutions are typically 
for the classical $SU(3)$ Yang - 
Mills theory in contradiction to monopole that is an exceptional solution. 
This allows to conclude that the confining properties of the classical 
$SU(3)$ Yang - Mills theory are general properties of this theory.
\end{abstract}
\narrowtext

\section{Introduction}

In quantum chromodynamic some field configurations can be 
interesting 
for phenomenological bag theory being used for an explanation
of quarks confinement. These configurations have to satisfy to the 
$SU(3)$ Yang - Mills equations and to confine inside itself  
the quark. For this purpose the $SU(2)$ magnetic bags have been 
constructed 
in Ref's \cite{prot}, \cite{adl} and \cite{per} where it has been shown 
that they can contain the fermions in some compact surface.
In Ref. \cite{sw} it has been also shown that the quantum particles 
with gauge charge can be confined only inside the domain of the $SU(2)$ 
Yang - Mills field configuration with infinite energy.
Recently the spherically symmetric solutions for the $SU(2)$ 
Yang - Mills equations, which are analogous to the black hole 
gravitational configurations, are founded in Ref's \cite{lun}, 
\cite{singl1}.
\par
In this article I want to show that the classical solutions of the 
$SU(3)$ Yang - Mills theory can have the confining properties and 
some solutions can be considered as a single quark (color charge) 
and a string (flux tube) between two quarks. To do this we 
examine the spherically and cylindrically symmetric cases.

\section{Spherically symmetric case}

The ansatz for the $SU(3)$ gauge field we take as in \cite{gal}:
\begin{mathletters}
\label{1}
\begin{eqnarray}
A^a & = & \frac{2\varphi(r)}{Ir^2} \left( \lambda ^2 x - \lambda ^5 y 
      + \lambda ^7 z\right ) + \frac{1}{2}\lambda ^a
      \left( \lambda ^a _{ij} + \lambda ^a_{ji} \right ) 
      \frac{x^ix^j}{r^2} w(r),
\label{1a:1}\\
A^a_i & = & \left( \lambda ^a_{ij} - \lambda ^a_{ji} \right )
        \frac {x^j}{Ir^2} \left(f(r) - 1\right ) +
        \lambda ^a_{jk} \left (\epsilon _{ilj} x^k + 
	\epsilon _{ilk} x^j\right ) \frac{x^l}{r^3} v(r)
\label{1b:2},
\end{eqnarray}
\end{mathletters}
here $\lambda ^a$ are the Gell - Mann matrixes; $a=1,2,\ldots ,8$
is color index; Latin indexes $i,j,k,l=1,2,3$ are the space indexes; 
$I^2=-1$; $r, \theta, \varphi$ are the spherically coordinate system. 
Substituting Eq's (\ref{1}) in the Yang - Mills equations:
\begin{equation}
\frac{1}{\sqrt{-g}}\partial _{\mu} \left (\sqrt {-g} {F^{a\mu}}_{\nu}
\right ) + f^{abc} {F^{b\mu}}_{\nu} A^c_{\mu} = 0,
\label{2}
\end{equation}
we receive the following $SU(3)$ equations system for 
$f(r), v(r), w(r)$ and $\varphi (r)$ functions:
\begin{mathletters}
\label{3}
\begin{eqnarray}
r^2f''& =& f^3 - f + 7fv^2 + 2vw\varphi - f\left (w^2 + \varphi ^2\right ),
\label{3a:1}\\
r^2v''& = & v^3 - v + 7vf^2 + 2fw\varphi - v\left (w^2 + \varphi ^2\right ),
\label{3b:2}\\
r^2w''& = & 6w\left (f^2 + v^2\right ) - 12fv\varphi,
\label{3c:3}\\
r^2\varphi''& = & 2\varphi\left (f^2 + v^2\right ) - 4fvw.
\label{3d:4}
\end{eqnarray}
\end{mathletters}
This set of equations is very difficult even for numerical
investigations. We will investigate a more simpler case when only
two functions are nonzero. It is easy to see that there can be only three 
cases. The first case is well-known monopole case by $(f,w=0)$
or $(v,w=0)$. Let us examine the two another interesting cases.

\subsection{The $SU(3)$ bag}

In this case we have $w=\varphi=0$ condition. Then, the set
(\ref{3}) has the following view:
\begin{mathletters}
\label{4}
\begin{eqnarray}
r^2f''& = & f^3 - f + 7fv^2,
\label{4c:3}\\
r^2v''& = & v^3 - v + 7vf^2.
\label{4d:4}
\end{eqnarray}
\end{mathletters}
We seek the solutions which are regular at origin and infinity.
This means that the solution at origin $r=0$ can be expanded
into a series:
\begin{mathletters}
\label{5}
\begin{eqnarray}
f & = & 1 + f_2 \frac{r^2}{2!} + \ldots ,
\label{5a:1}\\
v & = & v_3 \frac{r^3}{3!} + \ldots .
\label{5b:2}
\end{eqnarray}
\end{mathletters}
After substitution of Eq's (\ref{5}) into Eq's (\ref{4}) we receive 
that $f_2$ and $v_3$ coefficients are arbitrary. On the infinity 
we seek solutions as a following series:
\begin{mathletters}
\label{6}
\begin{eqnarray}
f & = & 1 + \frac{f_{-1}}{r} + \ldots ,
\label{6a:1}\\
v & = & \frac{v_{-2}}{r^2} + \ldots .
\label{6b:2}
\end{eqnarray}
\end{mathletters}
Analogously, the substitution of Eq's (\ref{6}) into Eq's(\ref{4}) shows that
$f_{-1}$ and $v_{-2}$ coefficients are also arbitrary.
\par
The numerical integration of the set (\ref{4}) with initial date (\ref{5}) 
shows that the solution is singular $(f\to \infty ,v\to \infty )$ by some 
$r=r_1$. In this case the approximate analytical solution near the 
singularity has the following form:
\begin{equation}
v \approx f \approx \frac{A}{r_1 - r},
\label{7}
\end{equation}
here $r<r_1$ and $A>0$ is some constant.
\par
The numerical integration of the set (\ref{4}) with initial data (\ref{6}) 
shows that here, too, the point $r=r_2$ exists in which the solution is 
singular: $(f\to \infty , v\to \infty )$. In this case the approximate 
analytical solution near the $r_2$ point has the following view:
\begin{equation}
v \approx f \approx \frac{B}{r - r_2},
\label{8}
\end{equation}
here $r>r_2$ and $B>0$ is also some constant.
\par
Selecting coefficients $f_2$ and $v_3$ (or $f_{-1}$ and $v_{-2}$), one can 
achieve the execution of $r_1=r_2$ condition. The according solution 
of the set (\ref{4}) (function $v(r)$) 
is presented on Fig.1 by $f_2=0.2; v_3=0.6; f_{-1}=2.07; 
v_{-2}\approx 3.8; r_1=r_2\approx 2.75; f(0)=f(\infty )=1; 
v(0)=v(\infty)=0$. The graph of $f(r)$ function is practically the same.
This solution is a bag with singular walls by $r=r_1=r_2$. 
\par
The similar solution for the $SU(2)$ gauge field is received in
\cite{singl1}.

\subsection{The $SU(3)$ bunker}

Here we examine $f=\varphi=0$ case. The case $v=\varphi=0$ is
analogous. Now the input equations have the following form:
\begin{mathletters}
\label{9}
\begin{eqnarray}
r^2 v''& = & v^3 - v - vw^2,
\label{9:1}\\
r^2 w''& = & 6w v^2.
\label{9:2}
\end{eqnarray}
\end{mathletters}
We seek the regular solution near $r=0$ point. The Eq's (\ref{9}) 
demand that $v$ and $w$ functions have the following view at origin
$r=0$:
\begin{mathletters}
\label{10}
\begin{eqnarray}
v & = & 1 + v_2 \frac{r^2}{2!} + \ldots,
\label{10:1}\\
w & = & w_3\frac{r^3}{3!} + \ldots .
\label{10:2}
\end{eqnarray}
\end{mathletters}
The numerical integration of Eq's (\ref{9}) is displayed on Fig.2,3. 
The asymptotical behaviour of received solution $(r\to \infty )$
is as follows:
\begin{mathletters}
\label{11}
\begin{eqnarray}
v & \approx & a \sin \left (x^{1+\alpha } + \phi _0\right ),
\label{11:1}\\
w & \approx & \pm\left [ (1 + \alpha ) x^{1 + \alpha } + 
\frac{\alpha}{4}\frac{\cos {\left (2x^{1 + \alpha} + 2\phi _0 \right )}}
{x^{1 + \alpha}}\right ],
\label{11:2}\\
3a^2 & = & \alpha(\alpha + 1).
\label{11:3}
\end{eqnarray}
\end{mathletters}
here $x=r/r_0$ is dimensionless radius; $r_0, \phi _0$ are some constants;
$\alpha \approx 0.37$.
For our potential $A^a_{\mu}$ we have the following nonzero color 
``magnetic'' and ``electric'' fields:
\begin{mathletters}
\label{12}
\begin{eqnarray}
H^a_{\varphi} & \propto & v',
\label{12:1}\\
H^a_{\theta} & \propto & v',
\label{12:2}\\
E^a_r & \propto & \frac{rw' - w}{r^2},
\label{12:3}\\
E^a_{\varphi} & \propto & \frac{vw}{r},
\label{12:4}\\
E^a_{\theta} & \propto & \frac{vw}{r},
\label{12:5}\\
H^a_r & \propto & \frac{v^2 - 1}{r^2},
\label{12:6}
\end{eqnarray}
\end{mathletters}
here for Eq's (\ref{12:1}), (\ref{12:2}) and (\ref{12:3}) the color index 
$a=1,3,4,6,8$ and for Eq's (\ref{12:4}), (\ref{12:5}) and (\ref{12:6}) 
$a=2,5,7$. Analyzing the asymptotical behaviour of the 
$H^a_{\varphi}, H^a_{\theta}, H^a_r$ and $E^a_{\varphi}, E^a_{\theta}$ 
fields we see that they are the strongly oscillating fields.
It is interesting that the radial components of the ``magnetic''
and ``electric'' fields drop to zero variously at infinity:
\begin{mathletters}
\label{13}
\begin{eqnarray}
H^a_r & \approx & \frac{1}{r^2},
\label{13:1}\\
E^a_r & \approx & \frac{1}{r^{1-\alpha}}.
\label{13:2}
\end{eqnarray}
\end{mathletters}
Among all the (\ref{12}) fields only the radial components of 
``electric'' fields are nonoscillating. From Eq's (\ref{11}) we see
that our solution has the oscillating part (\ref{11:1}) and confining
potential (\ref{11:2}). It is necessary to mark that obtained solution
is solution with arbitrary initial condition, whereas the monopole
solution is a special case of the initial condition. This allows to 
make a conclusion that the general spherically symmetric solution in the 
classical Yang - Mills theory has the confining properties. Although,
at the same time there are the oscillating potential, ``electric''
and ``magnetic'' fields.
The expression for an energy density has the following view:
\begin{equation}
\epsilon \propto 4\frac{v'^2}{r^2} + \frac{2}{3}\left (
\frac{w'}{r} - \frac{w}{r^2}\right ) + 4\frac{v^2w^2}{r^4} +
\frac{2}{r^2} \left (v^2 - 1\right ).
\label{14}
\end{equation}
This function is displayed on the Fig.3.
\par
We note the asymptotical form of the energy density followed
from (\ref{12}) condition:
\begin{equation}
\epsilon \approx \frac{2}{3} \frac{\alpha (1 + \alpha)^2 (3\alpha + 2)}
{x^{2-2\alpha }}.
\label{15}
\end{equation}
In the first approximation $\epsilon$ is nonoscillating on the infinity.
The asymptotical form of $\epsilon$ leads to that the energy of
such solution is infinity.
\par
It is very interesting what happens after quantization. It is very 
possible that such procedure smooths these strongly oscillating 
$SU(3)$ gauge field. In this case we will have only nonoscillating
confining potential (\ref{11:2}) and moreover, also nonoscillating 
energy density defined according to (\ref{15}).
\par
What is the physical meaning of this solution? It is possible that 
it is analogous to the Coulomb potential in electrostatics. 
But an electron can exist in empty space while a quark is not
observable in a free state. Therefore, the obtained solution can  
describe the color charge - ``quark''. 
The quotation marks indicate that we examine the simplified 
Eq's (\ref{4}) instead of complete Eq's (\ref{3}).
We emphasize that these two
solutions have the fundamental difference among themselves. The electron
has a singularity at origin by $r=0$, but  the ``quark'' at infinity
by $r=\infty$. In the second case such field configuration has the 
confining properties. It should be also noted that this solution has the 
asymptotical freedom property since at origin $r=0$ the gauge potential 
$A^a_{\mu} \to 0$.

\section{The gauge ``string''}

Let us write down the potential as follows:
\begin{mathletters}
\label{16}
\begin{eqnarray}
A^2_t & = & f(\rho ),
\label{16:1}\\
A^5_z & = & v(\rho ),
\label{16:2}\\
A^7_{\varphi} & = & \rho w(\rho ),
\label{16:3}
\end{eqnarray}
\end{mathletters}
here we introduce the cylindrical coordinate system $z, \rho , \varphi$.
The color index $a=2,5,7$ corresponds to the chosen $SU(2)\subset SU(3)$ 
inclusion. The Yang - Mills equations for potential (\ref{16}) are:
\begin{mathletters}
\label{17}
\begin{eqnarray}
f'' + \frac{f'}{\rho} & = & f\left (v^2 + w^2 \right ),
\label{17:1}\\
v'' + \frac{v'}{\rho} & = & v\left (-f^2 + w^2 \right ),
\label{17:2}\\
w'' + \frac{w'}{\rho}  - \frac{w}{\rho ^2}& = & w
\left (-f^2 + v^2 \right ),
\label{17:3}
\end{eqnarray}
\end{mathletters}
Let us examine the simplest case $w=0$. In this case equation set is:
\begin{mathletters}
\label{18}
\begin{eqnarray}
f'' + \frac{f'}{\rho} & = & fv^2 ,
\label{18:1}\\
v'' + \frac{v'}{\rho} & = & -vf^2 .
\label{18:2}
\end{eqnarray}
\end{mathletters}
At origin $\rho =0$ the solution has the following form:
\begin{mathletters}
\label{19}
\begin{eqnarray}
f & = & f_0 + f_2\frac{\rho ^2}{2} + \ldots ,
\label{19:1}\\
v & = & v_0 + v_2\frac{\rho ^2}{2} + \ldots .
\label{19:2}
\end{eqnarray}
\end{mathletters}
Substituting Eq's(\ref{19}) into (\ref{18}) system we find that:
\begin{mathletters}
\label{20}
\begin{eqnarray}
f_2 & = & \frac{1}{2}f_0v_0^2 ,
\label{20:1}\\
v_2 & = & -\frac{1}{2}v_0f_0^2 .
\label{20:2}
\end{eqnarray}
\end{mathletters}
The numerical integration is shown on Fig.2,4. It can be also shown that 
asymptotical behaviour of $f,v$ function and energy density $\epsilon$ is:
\begin{mathletters}
\label{21}
\begin{eqnarray}
f & \approx & 2\left [x + \frac{\cos \left (2x^2 + 2\phi _1\right )}
{16x^3} \right ] ,
\label{21:1}\\
v & \approx & \sqrt{2} \frac{\sin \left (x^2 + \phi _1 \right )}{x} ,
\label{21:2}\\
\epsilon & \propto & f'^2 + v'^2 + f^2v^2 \approx const,
\label{21:3}
\end{eqnarray}
\end{mathletters}
here $x=\rho /\rho _0$ is dimensionless radius; $\rho _0, \phi _1$ 
are  some constants. 
Again we have the confining potential $A^2_t=f(r)$ and strongly oscillating 
potential $A^5_z=v(r)$. Thus, this solution is a hollow or hump 
(in accordance with relation between $v_0$ and $f_0$ value) on the 
background of the constant energy density. On account of its cylindrical 
symmetry we can call this solution as a ``string''. The quotation marks 
indicate that this is the string from energetically point of view, not from 
potential $A^a_{\mu}$ and field $F^a_{\mu\nu}$ point of view. It can be 
expected that after quantization the oscillating functions will vanish 
and only confining potential and constant energy density will stay.
\par
It can be supposed that the obtained solution describes the classical gauge 
field between two ``quarks''. This must be verified using the lattice 
calculations (nonperturbative quantization). 

\section{Discussion}

Thus, we receive the classical solutions of the $SU(3)$ Yang - Mills 
theory. These solutions show us that the confinement 
is a general property in the classical $SU(3)$ Yang - Mills theory.
They are the typical solutions for the (\ref{3}) set since 
$f_2, v_2, v_3, f_{-1}, v_{-2}$ constants are arbitrary. 
In contradiction of this the monopole solution
is the exceptional solution. It has the finite energy whereas the 
typical solutions (obtained here) have 
infinite potential at infinity, infinite energy and consequently 
the confining properties. The confining solutions of this kind have 
the strong oscillating component of gauge potential, ``electric'' and
``magnetic'' fields. But it can be expected that these are only the 
classical properties vanishing in quantum theory.
\par
It can be supposed that the received here solutions have the physical
significance: The spherically symmetric case describes 
either a bag confining the quantum test particle or a classical 
single ``quark''; the cylindrically symmetric 
case describes a classical field distribution between two ``quarks''.
\par
In second and third  cases, the situation greatly differs from that what 
happens with electron. An isolated electron exists in nature and it 
generates the electric field decreases at infinity. An isolated quark 
does not exist in nature. It is possible that it forms the  
confining field distribution. The two 
interacted electrons generate electric field appearing as a superposition 
of electric field from two electrons. The two quarks generate the string. 
It can be supposed that the gauge ``string'' obtained above is a classical
model of such field distribution. It appears as a string on the 
background of the field with constant energy density. Just the same
it should be remarked that here we have also the strong oscillating 
fields. These fields maybe will be excluded by quantization.

\newpage
\centerline{List of figure captions}
Fig.1. $v(r)$ function for the $SU(3)$ bag.
\par
Fig.2. The confining potentials. $1$ is the $w(r)$ function for 
the $SU(3)$ bunker, $2$ is the $f(\rho )$ function for the $SU(3)$ 
``string''.
\par
Fig.3. The $SU(3)$ bunker. $1$ is the oscillating potential $v(r)$
and $2$ is the energy density $\epsilon (r)$.
\par
Fig.4. The $SU(3)$ ``string''. $1$ is the oscillating potential
$v(\rho )$ and $2$ is the energy density $\epsilon (\rho )$.
\end{document}